\documentclass[twocolumn,showpacs,prd]{revtex4}
\usepackage{mathrsfs,bm,multirow}
\usepackage{longtable,lscape}
\usepackage{txfonts}
\usepackage{amssymb}
\usepackage{indentfirst}
\usepackage{graphicx,,booktabs}
\usepackage{color}
\usepackage{amssymb}

\begin{document}

\title{Predictions of charged charmonium-like structures with hidden-charm and open-strange channels}
\author{Dian-Yong Chen$^{1,3}$}
\email{chendy@impcas.ac.cn}
\author{Xiang Liu$^{1,2}$\footnote{Corresponding author}}\email{xiangliu@lzu.edu.cn}
\author{Takayuki Matsuki$^4$}
\email{matsuki@tokyo-kasei.ac.jp}
\affiliation{$^1$Research Center for Hadron and CSR Physics,
Lanzhou University $\&$ Institute of Modern Physics of CAS,
Lanzhou 730000, China\\
$^2$School of Physical Science and Technology, Lanzhou University,
Lanzhou 730000, China\\
$^3$Nuclear Theory Group, Institute of Modern Physics, Chinese
Academy of Sciences, Lanzhou 730000, China\\
$^4$Tokyo Kasei University, 1-18-1 Kaga, Itabashi, Tokyo 173-8602,
Japan}

\begin{abstract}

We propose the initial single chiral particle emission mechanism,
with which the hidden-charm di-kaon decays of higher charmonia and
charmonium-like states are studied. Calculating the distributions
of differential decay width, we obtain the line shape of the
$J/\psi K^+$ invariant mass spectrum of $\psi_i\to J/\psi K^+K^-$,
where $\psi_i=\psi(4415), Y(4660)$, and $\psi(4790)$. Our
numerical results show that there exist enhancement structures
with both hidden-charm and open-strange decays, which are near the
$D\bar{D}_s^*/D^*\bar{D}_s$ and $D^*\bar{D}_s^*/\bar{D}^*{D}_s^*$
thresholds. These charged charmonium-like structures predicted in
this Letter can be accessible in future experiments, especially
BESIII, BelleII, and SuperB.

\end{abstract}
 \maketitle

It is exciting to notice that the BESIII Collaboration has, for
the first time, reported the observation of a charged
charmonium-like structure $Z_c(3900)$ in $e^+e^-\to
J/\psi\pi^+\pi^-$ at $\sqrt{s}=4260$ MeV \cite{BESnew}. As an
enhancement structure around 3.9 GeV with statistical significance
$9.0\sigma$, $Z_c(3900)$ exists in the $J/\psi\pi^\pm$ invariant
mass spectrum \cite{BESnew}. This observation confirms one of our
predictions by the {\it{Initial Single Pion Emission}} (ISPE)
mechanism \cite{Chen:2011xk}, where we indicated that a charged
charmonium-like structure near the $D\bar{D}^*$ threshold is
observable in the $Y(4260)\to J/\psi\pi^+\pi^-$ process (see Ref.
\cite{Chen:2011xk} for details). In addition, we also need to
indicate that the study of the exotic molecular states composed of
$D^{(*)}$ and $\bar{D}^{*}$ and the relevant prediction of the
isovector $D\bar{D}^*$ molecular state was given in Ref.
\cite{Sun:2011uh} before BESIII's observation \cite{BESnew}.

The ISPE mechanism was first proposed to explain Belle's
observations of two charged bottomonium-like structures
$Z_b(10610)$ and $Z_b(10650)$ appearing in the hidden-bottom
dipion decays of $\Upsilon(5S)$ \cite{Collaboration:2011gja}. Via
the ISPE mechanism, two $Z_{b}$ structures near the $B\bar{B}^{*}$
and $B^*\bar{B}^*$ thresholds can be understood
\cite{Chen:2011pv}. In addition, our results also naturally answer
the question of why the charged bottomonium-like structures near
the $B\bar{B}$ threshold was not found by Belle. In Ref.
\cite{Chen:2011pv}, we further indicated that the ISPE mechanism
can be applied to the hidden-charm dipion decays of the higher
charmonia $\psi(4040)$, $\psi(4160)$, $\psi(4415)$, and
charmonium-like state $Y(4260)$, which is due to the similarity
between charmonium and bottomonium processes. Later, by studying
the line shapes of the differential decay widths for $\psi(4040)$,
$\psi(4160)$, $\psi(4415)$, and $Y(4260)$ decays into
$J/\psi\pi^+\pi^-$, $\psi(2S)\pi^+\pi^-$, and $h_c(1P)\pi^+\pi^-$,
we predicted the sharp peak structures close to the $D\bar{D}^*$
and $D^*\bar{D}^*$ thresholds in the corresponding $J/\psi\pi^+$,
$\psi(2S)\pi^+$, and $h_c(1P)\pi^+$ invariant mass spectra
\cite{Chen:2011xk}. Thus, the charged charmoniu-like structure
$Z_c(3900)$ newly observed by BESIII can be an important test for
the ISPE mechanism.

Besides these theoretical predictions relevant to the hidden-charm
dipion decays of higher charmonia, via the ISPE mechanism we have
also predicted the charged bottomonium-like structures near the
$B\bar{B}^*$ and $B^*\bar{B}^*$ thresholds in the hidden-bottom
dipion decays of $\Upsilon(11020)$ \cite{Chen:2011pu}, and two
charged strangeonium-like structures observable in the $Y(2175)
\to\phi(1020)\pi^{+} \pi^{-}$ process \cite{Chen:2011cj}. We expect
more experimental progress to test these predictions of ours
presented in Refs. \cite{Chen:2011xk,Chen:2011pu,Chen:2011cj}.

Indeed this new experimental observation given by BESIII
\cite{BESnew} not only brings us excitement, but also further
stimulates us to think about how to incorporate the ISPE mechanism
with other phenomena, i.e., its application to more decay
processes, and make more theoretical predictions.

Along the way, we notice that both of the pion and kaon can be
categorized as chiral particles. Thus, the hidden-charm di-kaon
decays of a higher charmonium are the intriguing processes, and
the {\it initial single chiral particle emission} (ISChE)
mechanism can play an important role in these decays, which is a
natural extension of the ISPE mechanism. To some extent, the ISPE
mechanism can be included in the ISChE mechanism proposed in this
work.

\begin{figure}[htp]
\begin{tabular}{cc}
\includegraphics[width=0.18\textwidth]{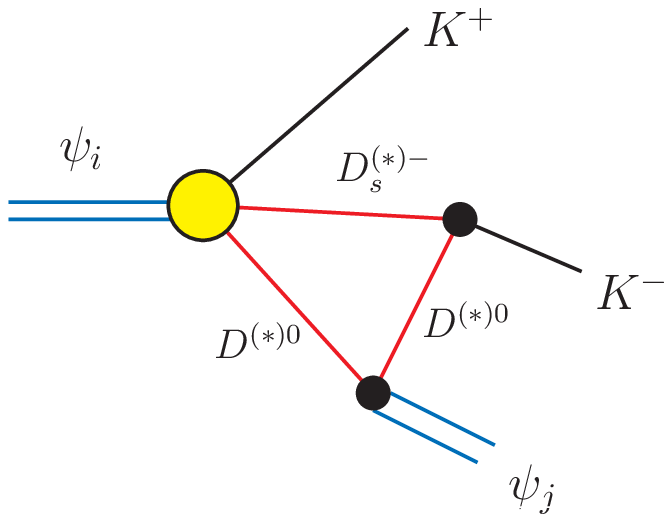} &
\includegraphics[width=0.18\textwidth]{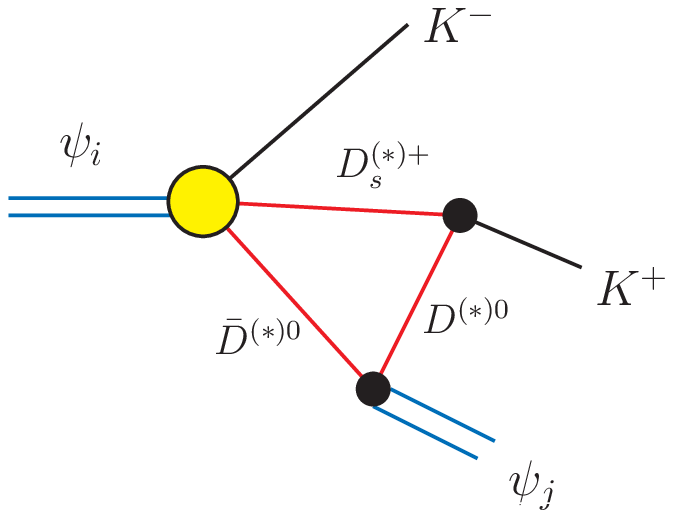} \\
$(a)$&$(b)$
\end{tabular}
\caption{(color online). Typical hadron-level diagrams of
hidden-charm di-kaon decay of higher charmonium by the ISChE
mechanism. Here, $\psi_{i}$ denotes the initial higher charmonium
and $\psi_{j}$ is a charmonium in the final state.
}\label{Fig:typical}
\end{figure}

In the following, we will specify the concrete process to
illustrate the physical picture of the ISChE mechanism. Under the
ISChE mechanism, the hidden-charm di-kaon decay of a higher
charmonium occurs via the triangle loop constructed by charmed and
charmed-strange mesons, which are shown in Fig. \ref{Fig:typical}.
The kaon with a continuous energy distribution emitted by an
initial higher charmonium makes the intermediate charmed and
charmed-strange mesons with low momenta easily interact with each
other and then transform into a charmonium and a kaon. As
non-perturbative QCD effects, the ISChE mechanism can be, in a word, 
the coupled-channel effect.

Applying the ISChE mechanism to the processes shown in Fig.
\ref{Fig:typical}, one requires two relations
$m_{\psi_i}>m_{K}+m_{D^{(*)}}+m_{D_{s}^{(*)}}$ and
$m_{\psi_i}>m_{\psi_j}+m_{K^+}+m_{K^-}$, which provide the
criteria to choose suitable processes in Fig. \ref{Fig:typical}.
The minimum of the sum of masses of the kaon, $D^{(*)}$, and
$D_{s}^{(*)}$ is about 4331 MeV. Thus, the mass of an initial
higher charmonium must be higher than 4331 MeV, when we study the
hidden-charm di-kaon decays of a higher charmonium by the ISChE
mechanism. Thus, in this work the hidden-charm di-kaon decays of
higher charmonia or charmonium-like states include
\begin{eqnarray}
&&\psi(4415)\to \bar{K}[D_s\bar{D}]/{K}[\bar{D}_sD]\to J/\psi K^+K^-,\label{h1}\\
&&Y(4660)\to \bar{K}[D_s^{(*)}\bar{D}^{(*)}]/{K}[\bar{D}_s^{(*)}{D}^{(*)}]\to J/\psi K^+K^-,\label{h2}\\
&&\psi(4790)\to \bar{K}[D_s^{(*)}\bar{D}^{(*)}]/{K}
[\bar{D}_s^{(*)}{D}^{(*)}]\to J/\psi K^+K^-\label{h3},
\end{eqnarray}
where $\psi(4415)$ is the first charmonium with mass above 4331
MeV listed by the Particle Data Group  \cite{Beringer:1900zz}.
$Y(4660)$ is a charmonium-like state reported by Belle
\cite{Wang:2007ea} and recently confirmed by BaBar
\cite{Lees:2012pv}, which was reported in studying $e^{+}e^{-}\to
\psi(2S)\pi^{+}\pi^{-}$ process. In addition, $\psi(4790)$ was
predicted as $n^{2s+1}L_J=5^3S_1$ charmonium by using the
resonance spectrum expansion model and analyzing the experimental
data \cite{vanBeveren:2008rt}. Studying these hidden-charm di-kaon
decays of $\psi(4415)$, $Y(4660)$, and $\psi(4790)$, we will
answer the question of whether there exist charged charmonium-like
structures with both of hidden-charm and open-strange decays.

Via the ISChE mechanism, the hidden-charm di-kaon decays of
$\psi(4415)$, $Y(4660)$, and $\psi(4790)$ can be depicted by the
schematic diagrams shown in Fig. \ref{Fig:typical}. When calculating
these diagrams, we adopt the effective Lagrangian approach, where
the Lagrangians describing the interaction vertices are
\cite{Oh:2000qr, Casalbuoni:1996pg,
Colangelo:2002mj,Colangelo:2003sa}
\begin{eqnarray}
\mathcal{L}_{\psi^\prime \mathcal{D}^{(\ast)} \mathcal{ D}^{(\ast)}
\mathcal{ P}} &=& -ig_{\psi^\prime \mathcal{DDP}} \varepsilon_{\mu
\nu \alpha \beta} \psi^{\prime \mu} \partial^\nu \mathcal{D}
\partial^{\alpha} \mathcal{P} \partial^\beta
\mathcal{D}^\dagger \nonumber\\
&&+ g_{\psi^\prime \mathcal{D}\mathcal{D}^\ast \mathcal{P}}
\psi^{\prime \mu} \left( \mathcal{D} \mathcal{P} \mathcal{D}^{\ast
\dagger}_\mu + \mathcal{D}^\ast_\mu \mathcal{P}
\mathcal{D}^{\dagger}\right)
\nonumber\\
&&- ig_{\psi^\prime \mathcal{D}^{\ast} \mathcal{D}^\ast
\mathcal{P}} \varepsilon_{\mu \nu \alpha \beta} \psi^{\prime \mu}
\mathcal{D}^{\ast \nu}
\partial^\alpha \mathcal{P} \mathcal{D}^{\ast \beta \dagger}
\nonumber\\
&&-ih_{\psi^\prime \mathcal{D}^\ast \mathcal{D}^\ast \mathcal{P}}
\varepsilon_{\mu \nu \alpha \beta} \partial^\mu \psi^{ \prime\nu}
\mathcal{D}^{\ast \alpha}\mathcal{P} \mathcal{D}^{\ast
\beta\dagger}, \label{Eq:Lag1}\\
\mathcal{L}_{\psi \mathcal{D}^{(\ast)} \mathcal{D}^{(\ast)}} &= &
ig_{\psi \mathcal{DD}} \psi^\mu \left(\partial_\mu \mathcal{D}
\mathcal{D}^{\dagger} - \mathcal{D} \partial_\mu \mathcal{D}^\dagger
\right) \nonumber\\
&&-g_{\psi \mathcal{D}^\ast \mathcal{D}} \varepsilon^{\mu \nu
\alpha \beta} \partial_{\mu} \psi_\nu \left( \partial_\alpha
\mathcal{D}^\ast_\beta \mathcal{D}^{\dagger} + \mathcal{D}
\partial_{\alpha} \mathcal{D}^{\ast \dagger}_{\beta}\right)\nonumber\\
&&-ig_{\psi \mathcal{D}^\ast \mathcal{D}^\ast } \left\{ \psi^\mu
\left( \partial_\mu \mathcal{D}^{\ast \nu} \mathcal{D}^{\ast
\dagger}_{\nu}- \mathcal{D}^{\ast \nu} \partial_\mu
\mathcal{D}^{\ast \dagger}_{\nu}\right)\right.\nonumber\\
&& + \left. \left( \partial_{\mu} \psi_\nu \mathcal{D}^{\ast \nu}
-\psi_{\nu} \partial_{\mu} \mathcal{D}^{\ast
\nu}\right) \mathcal{D}^{\ast \mu\dagger} \right.\nonumber\\
&& + \left. \mathcal{D}^{\ast \mu} \left( \psi^\nu
\partial_\mu \mathcal{D}^{\ast\dagger}_{\nu} -\partial_{\mu}
\psi_\nu \mathcal{D}^{\ast \nu \dagger }\right)\right\},  \label{Eq:Lag2}\\
\mathcal{L}_{\mathcal{D}^{(\ast)}\mathcal{D}^{(\ast)} \mathcal{P}}
&=& -ig_{\mathcal{D}^\ast \mathcal{D}\mathcal{P}} \left(\mathcal{D}
\partial_\mu \mathcal{P} \mathcal{D}^{\ast \mu \dagger}
-\mathcal{D}^{\ast \mu} \partial_\mu \mathcal{P}
\mathcal{D}^{\dagger}\right) \nonumber\\
&&-g_{\mathcal{D}^\ast \mathcal{D}^\ast \mathcal{P}}
\varepsilon^{\mu \nu \alpha \beta} \partial_{\mu}
\mathcal{D}^{\ast}_\nu \mathcal{P} \partial_\alpha
\mathcal{D}_\beta^{\ast \dagger}, \label{Eq:Lag3}
\end{eqnarray}
where $\mathcal{D}^{(\ast)}= \left(D^{(\ast)0}, D^{(\ast)+},
D_s^{(\ast)+} \right)$. $\mathcal{P}$ is a pseudoscalar meson matrix.

\begin{table}
\caption{{Optimal values of the coupling constants involved in the
present work. These coupling constants can be related to the gauge
coupling $g$ by $g_{\psi D D} = g_{\psi D^\ast D^\ast}
{m_{D^\ast}}/{m_D} = g_{\psi D^\ast D } m_{\psi}
\sqrt{{m_D}/{m_D^\ast}} ={m_{\psi}}/{f_\psi}$, $g_{\psi
D_s^{(\ast)}D_s ^{(\ast)}} =\sqrt{(m_{D_s}^{(\ast)}
m_{D_s}^{(\ast)})/(m_D^{(\ast)} m_D^{(\ast)})} g_{\psi D D}$,
$g_{D_s^\ast D^\ast K} = \sqrt{m_{D_s}^\ast/ m_{D^\ast} }
{2g}/{f_\pi}$, and $ {g_{D_s^\ast D K}}/{\sqrt{m_{D_s^\ast m_D }}}
= {g_{D^\ast D_s K}}/{\sqrt{m_{D^\ast m_{D_s} }}} ={2g}/{f_\pi}$
\cite{Colangelo:2003sa}, where $f_{\psi}$ is the decay constants
of $J/\psi$ ,which is evaluated from the leptonic decay width of
$J/\psi$. With $\Gamma_{J/\psi \to e^+ e^-}=5.55$ keV
\cite{Beringer:1900zz}, we have $f_{\psi}=416$ MeV. $f_\pi=132$
MeV is the pion decay constant and $g=0.59$ is estimated from the
partial decay width of $D^\ast \to D\pi$ \cite{Beringer:1900zz}.
\label{Tab:Couple}}}
\begin{tabular}{cccccc}
\toprule[1pt] %
Coupling & Value & Coupling & Value & Coupling & Value\\
\midrule[1pt] %
 $g_{\psi DD}$                 & $7.44$                     &
 $g_{\psi D^\ast D}$           & $2.49\ \mathrm{GeV}^{-1}$  &
 $g_{\psi D^\ast D^\ast}$      & $8.01$                      \\
 $g_{\psi D_sD_s}$             & $7.84$                     &
 $g_{\psi D_s^\ast D_s}$       & $2.62\ \mathrm{GeV}^{-1}$  &
 $g_{\psi D_s^\ast D_s^\ast}$  & $8.42$                      \\
 $g_{D_s^\ast D K}$            & $17.76$                    &
 $g_{D^\ast D_s K}$            & $17.78$                    &
 $g_{D_s^\ast D^\ast K }$      & $9.16\ \mathrm{GeV}^{-1}$   \\
\bottomrule[1pt]%
\end{tabular}
\end{table}

In the heavy quark limit and using chiral symmetry, the coupling
constants can be connected to one gauge coupling; {the definite
values of all the coupling constants are listed in Table
\ref{Tab:Couple}.} In the present work, the initial vector
charmonia are above the threshold of $D^{(\ast)} D_{s}^{(\ast)}
K$, thus the coupling constants of $\psi^\prime
\mathcal{D}^{(\ast)} \mathcal{D}^{(\ast)} K$ should be evaluated
from the corresponding partial decay widths. In addition, because
we mainly concern ourselves with the line shape of the $J/\psi K$
invariant mass spectrum in the frame of the ISChE mechanism, the
interferences between different intermediate states are not taken
into account in the present calculations. Here, we assume $g_{\psi
\mathcal{D}^\ast \mathcal{D}^\ast \mathcal{P}}=h_{\psi
\mathcal{D}^\ast \mathcal{D}^\ast \mathcal{P}}$, which results
from the heavy quark limit and $SU(4)$ symmetry \cite{Oh:2000qr}.

With the effective Lagrangians listed in Eqs.
(\ref{Eq:Lag1})-(\ref{Eq:Lag3}), one can obtain the general form
of the amplitudes of $\psi_i (p_0) \to K^+(p_3)
[\bar{D}_s^{(\ast)} (p_1) D^{(\ast)}(p_2) ] \to K^+(p_3) K^-(p_4)
\psi_j(p_5) $ and $\psi_i (p_0) \to K^-(p_4) [D^{(\ast)}_s (p_1)
\bar{D}^{(\ast)}(p_2) ] \to K^+(p_3) K^-(p_4) \psi_j(p_5) $
corresponding to Figs. \ref{Fig:typical} (a) and \ref{Fig:typical}
(b), respectively, i.e.,
\begin{eqnarray}
&&\mathcal{M}\{\psi_i  \to K^+ [\bar{D}_s^{(\ast)} D^{(\ast)} ] \to
K^+ K^- \psi_j \} \nonumber\\
&=& \prod_i g_i \int \frac{d^4q}{(2\pi)^4}
\frac{[p_3,p_4,p_5,q]_{\mu \nu} \epsilon_{\psi_i}^\mu
\epsilon_{\psi_j}^\nu }{\big[(p_4+q)^2 -m^2_{D_s^{(\ast)}}\big]
\big[(p_5-q)^2 -m_{D^{(\ast)}}^2 \big]} \nonumber\\
&&\times \frac{1}{q^2-m_{D^{(\ast)}}^2}
\mathcal{F}(q^2,m^2_{D^{(\ast)}}), \\
&&\mathcal{M}\{\psi_i  \to K^- [{D}_s^{(\ast)} \bar{D}^{(\ast)}
] \to K^+ K^- \psi_j\} \nonumber\\
&=& \prod_i g_i \int \frac{d^4q}{(2\pi)^4}
\frac{[p_3,p_4,p_5,q]_{\mu \nu} \epsilon_{\psi_i}^\mu
\epsilon_{\psi_j}^\nu }{\big[(p_3+q)^2 -m^2_{D_s^{(\ast)}}\big]
\big[(p_5-q)^2 -m_{D^{(\ast)}}^2 \big]} \nonumber\\
&&\times \frac{1}{q^2-m_{D^{(\ast)}}^2}
\mathcal{F}(q^2,m^2_{D^{(\ast)}}),
\end{eqnarray}
where $[p_3,p_4,p_5,q]^{\mu \nu}$ denotes the tensor function of
$p_3$, $p_4$, $p_5$ and the integral momentum $q$, which can be
constructed by the effective Lagrangians in Eqs.
(\ref{Eq:Lag1})-(\ref{Eq:Lag3}). In order to describe the internal
structures of the exchange mesons and its off shell effects, we
introduce a form factor in the form $\mathcal{F}( q^2,m_E^2)
=(m_E^2- \Lambda^2)/(q^2 -\Lambda^2)$. The parameter $\Lambda$ is
parametrized as $\Lambda=\alpha \Lambda_{QCD} + m_E$ with
$\Lambda_{QCD}=0.22$ GeV. In Ref. \cite{Chen:2011cj}, we
numerically proved that the line shapes are weakly dependent on
the unique parameter $\alpha$ and hence $\alpha=1$ is used in the
present work.

In the following we use $\mathcal{M}_{AB}^{C}$ to indicate the
process where initial $\psi_i(p_0)$ decays into a meson pair
$A(p_1) B(p_2)$ with $K^+(p_3)$ emission and the meson pair
subsequently transits into $K^-(p_4) \psi_j(p_5)$ by exchanging a
meson $C(q)$. Here, we consider the ISChE mechanism with different
intermediate states and search for possible enhancement structures
near the threshold of $D_s^{(\ast)} D^{(\ast)}$ in the invariant
mass spectrum of $J/\psi K^+$. The total amplitudes from different
intermediate states in the frame of ISChE are
\begin{eqnarray}
\mathcal{A}^{\mathrm{tot}}_{D D_s}& =&  \mathcal{M}_{D
\bar{D}_s}^{D_s^\ast} + \mathcal{M}_{\bar{D}_s D }^{\bar{D}^\ast} +\dots ,\\
\mathcal{A}^{\mathrm{tot}}_{D^\ast D_s+ D D_s^\ast} &=&
\mathcal{M}_{D\bar{D}_s^\ast}^{D_s^\ast} +\mathcal{M}_{D^\ast
\bar{D}_s}^{D_s} + \mathcal{M}_{D^\ast \bar{D}_s}^{D_s^\ast}
+\mathcal{M}_{\bar{D}_s D^\ast}^{\bar{D}^\ast }\nonumber\\&&
+\mathcal{M}_{ \bar{D}_s^\ast D}^{\bar{D}}+
\mathcal{M}_{ \bar{D}_{s}^\ast D}^{\bar{ D}^\ast} + \dots, \\
\mathcal{A}_{D_s^\ast D^\ast}^{\mathrm{tot}} &=&
M_{D^\ast\bar{D}_s^\ast}^{D_s^\ast} +
M_{D^\ast\bar{D}_s^\ast}^{D_s^\ast} + \mathcal{M}_{\bar{D}_s^\ast
D^\ast}^{\bar{D}}+ \mathcal{M}_{\bar{D}_s^\ast
D^\ast}^{\bar{D}^\ast} + \dots,\nonumber\\
\end{eqnarray}
where '$\dots$' denotes the contributions from the ISChE mechanism with
$K^-$ emitted from initial charmonia.

\begin{figure}[htb]
\centering %
\scalebox{1}{\includegraphics{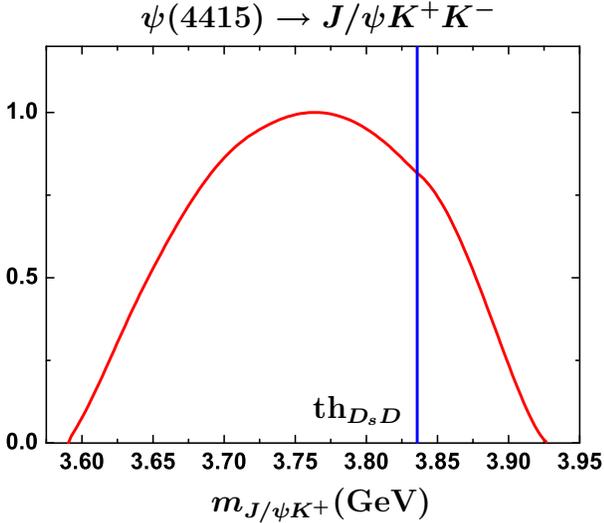}}%
\caption{(Color online). The line shape of $\psi(4415) \to J/\psi
K^+ K^-$ in the ISChE frame with $D_s \bar{D} +H.C.$ intermediate
state contributions. \label{Fig:psi4415}}
\end{figure}

\begin{figure*}[htbp]
\centering %
\scalebox{1.25}{\includegraphics{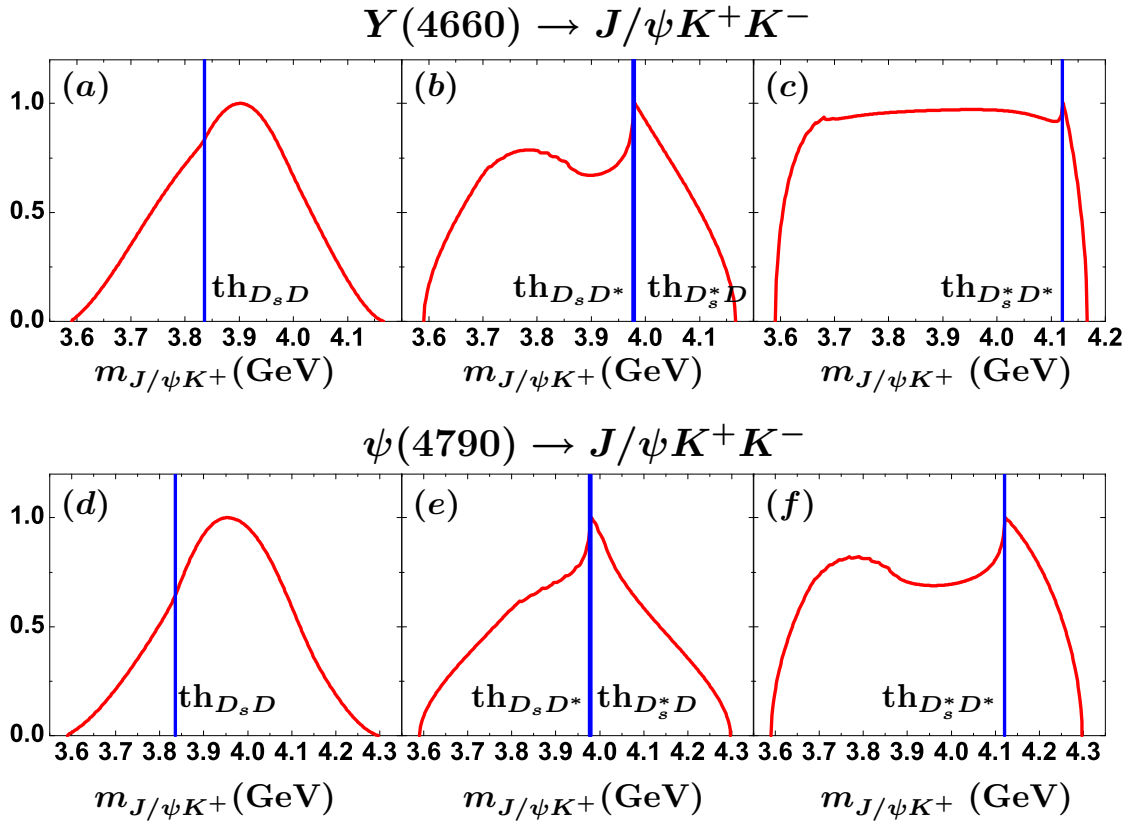}}%
\caption{(color online). Dependence of the distribution of
$d\Gamma/dm_{J/\psi K^+}$ on the $J/\psi K^+$ invariant mass
spectrum (red solid curves). The diagrams (a) and (d), the
diagrams (b) and (e), and the diagrams (c) and (f), are the
results considering the intermediate $D \bar{D}_s+H.C.$, $
{D}^{*}\bar{D}_s+D  \bar{D}_s^\ast +H.C.$, and $D^*\bar{D}_s^*
+H.C.$ contributions, respectively. Here, the line shape of the
distribution of $d\Gamma/dm_{J/\psi K^+}$ is normalized to 1.
\label{Fig:jpsikk}}
\end{figure*}

With the above amplitudes, we can obtain the invariant mass
distributions of $J/\psi K^+$ for $\psi_i \to J/\psi K^+ K^-$ in
the frame of the ISChE mechanism. In Figs. \ref{Fig:psi4415} and
\ref{Fig:jpsikk}, we show the numerical results of the
distribution of $d\Gamma/dm_{J/\psi K^+}$ dependent on the $J/\psi
K^+$ invariant mass spectrum. In our calculation, we consider the
intermediate$D \bar{D}_s+H.C.$, $ {D}^{*}\bar{D}_s+D
\bar{D}_s^\ast +H.C.$, and $D^*\bar{D}_s^* +H.C.$ contributions
separately, and the line shape of the distribution of
$d\Gamma/dm_{J/\psi K^+}$ is normalized to 1. The vertical lines
are the corresponding thresholds. Based on these theoretical
results, we find

1. The vector charmonium $\psi(4415)$ is below the thresholds of
$D_s^{\ast} \bar{D} \bar{K} / D_s \bar{D}^{\ast} \bar{K}$ and
$D_s^\ast \bar{D}^\ast \bar{K}$. Thus, under the ISChE mechanism,
$\psi(4415)\to J/\psi K^+K^-$ occurs only via the intermediate
$D\bar{D}_s+ H.C$. The $d\Gamma/dm_{J/\psi K^+}$ distribution is
presented in Fig. \ref{Fig:psi4415}. The line shape is smooth and
does not show a sharp peak near the $D\bar{D}_s+H.C.$ threshold,
which also holds for $Y(4660)\to J/\psi K^+K^-$ and $\psi(4791)\to
J/\psi K^+K^-$ decays (see Figs. \ref{Fig:jpsikk} (a) and
\ref{Fig:jpsikk} (d)).

2. If considering the intermediate $ {D}^{*}\bar{D}_s+D
\bar{D}_s^\ast +H.C.$ contribution to $Y(4660)\to J/\psi K^+K^-$,
we notice that there exist two structures in the line shape shown
in Fig. \ref{Fig:jpsikk} (b). The higher one is a sharp peak
structure near the $D_s\bar{D}^{*}/\bar{D}_s^{*}{D}$ thresholds
while the lower one is a broad structure as a reflection of the
higher one. If only the intermediate $D^*\bar{D}_s^*+H.C.$ is
included, the resultant distribution of $d\Gamma/dm_{J/\psi K^+}$
of $\psi(4660)\to J/\psi K^+K^-$ gives a small sharp peak
structure around the $D^*\bar{D}_s^*/\bar{D}^* D_s^*$ threshold.
In addition, its reflection raised as $3.7$ GeV is obscure (see
Fig. \ref{Fig:jpsikk} (c) for more details).

3. As for $\psi(4790)\to J/\psi K^+K^-$ decay, we present the
corresponding line shapes of the $d\Gamma/dm_{J/\psi K^+}$
distribution by considering the intermediate ${D}^{*}\bar{D}_s+D
\bar{D}_s^\ast +H.C.$ and $D^*\bar{D}_s^* +H.C.$ contributions in
the diagrams of Figs. \ref{Fig:jpsikk} (e) and (f). The Figure
\ref{Fig:jpsikk} (e) shows an enhancement structure existing in
the $d\Gamma/dm_{J/\psi K^+}$ distribution, which is composed of a
sharp peak near the $D_s\bar{D}^{*}/\bar{D}_s^{*}{D}$ threshold
and its reflection. This situation is different from that of
$Y(4660)\to J/\psi K^+K^-$ just mentioned above. Figure
\ref{Fig:jpsikk} (f) indicates a sharp peak structure close to the
$D^*\bar{D}_s^*/\bar{D}^* D_s^*$ threshold and a broad structure
corresponding to its reflection, which are more clear compared
with the structures shown in Fig. \ref{Fig:jpsikk} (c).

In this work, we have extended the ISPE mechanism and have
proposed the ISChE mechanism to study the hidden-charm di-kaon
decays of higher charmonia and charmonium-like state, where we
have chosen three typical processes $\psi(4415)\to J/\psi K^+K^-$,
$Y(4660)\to J/\psi K^+K^-$, and $\psi(4790)\to J/\psi K^+K^-$.
Under the ISChE mechanism, we have calculated their differential
decay widths and presented their dependence on the $J/\psi K^+$
invariant mass spectrum. Our results have shown that there do not
exist peak structures near $D\bar{D}_s/D_s\bar{D}$ for the decays
of concern. However, we have found sharp enhancement structures
around the $D_s^{(*)}\bar{D}^{*}/\bar{D}_s^{(*)}{D}^{*}$ and
$D^*\bar{D}_s^*/\bar{D}^* D_s^*$ thresholds, where the expected
enhancement structures in the $J/\psi K^+$ invariant mass spectra
have charge and are with both hidden-charm and open-strange decays
in the final state, which are very peculiar.

According to our study, we suggest future experiments to carry out
the search for these charged charmonium-like structures with
hidden-charm and open-strange decays. If our predictions are
confirmed in the future, the ISChE mechanism existing in the
hidden-charm di-kaon decays of higher charmonia or charmonium-like
state can be tested further, which will also show that the ISChE
mechanism can be a universal mechanism for heavy quarkonia. At
present, BESIII has already obtained beautiful experimental
results, the observation of a charged charmonium-like structure
near $D\bar{D}^*$ threshold in $e^+e^-\to J/\psi\pi^+\pi^-$ at
$\sqrt{s}=4260$ MeV. Among the hidden-charm di-kaon decays
discussed in this Letter, $\psi(4415)\to J/\psi K^+K^-$,
$Y(4660)\to J/\psi K^+K^-$ is accessible at BESIII. It will be a
good opportunities for BESIII to study charmonium-like structures
$XYZ$. In addition, the Belle and BaBar and forthcoming BelleII
and SuperB Collaborations will be good platforms to carry out the
search for these charged charmonium-like structures with
hidden-charm and open-strange decays.

\vfil
\section*{Acknowledgement}

X.L. would like to thank Shan Jin for a fruitful and inspired
discussion. This project is supported by the National Natural
Science Foundation of China under Grants No. 11222547, No.
11175073, No. 11005129 and No. 11035006, the Ministry of Education
of China (FANEDD under Grant No. 200924, SRFDP under Grant No.
20120211110002 , NCET, the Fundamental Research Funds for the
Central Universities), the Fok Ying Tung Education Foundation (No.
131006), and the West Doctoral Project of the Chinese Academy of
Sciences.

\end{document}